\documentclass[aps,prb,draft,groupedaddress]{revtex4}
\usepackage{graphicx}
\begin{document}
\title{Magnetic study of an amorphous conducting polyaniline}
\author{Debangshu Chaudhuri}
\author{Ashwani Kumar}
\author{D. D. Sarma$^{a)}$}
\affiliation{Solid State and Structural Chemistry Unit, Indian
Institute of Science, Bangalore-560012, India}
\author{M. Garc$\acute{i}$a-Hern$\acute{a}$ndez}
\affiliation{Instituto de Ciencia de Materiales de Madrid, CSIC, Cantoblanco, E-28049
Madrid, Spain}
\author{J. Joshi}
\author{S. V. Bhat}
\affiliation{Department of Physics, Indian Institute of Science, Bangalore 560~012, India}

\begin{abstract} We show that newly found BF$_3$-doped polyaniline, though
highly conducting, remains amorphous. Magnetic studies reveal many
unusual properties, while suggesting that the intrinsic
conductivity of this system is significantly larger than all other
known forms of conducting polyaniline, establishing it as an
interesting class of highly conducting amorphous polymer.
\end{abstract}
\maketitle

\newpage

Over a number of years, proton-doped polyaniline (PANI) has become
one of the most extensively studied conducting polymers owing to
its high electrical conductivity and consequent possibilities of
technological applications. Undoped polyaniline is amorphous.
However, proton doping induces the formation of highly conducting,
crystalline domains separated by amorphous, insulating
barriers.~\cite{hclSM87} It is reported that the doped charge
carriers in the protonated PANI are predominantly localized in the
amorphous regions, giving rise to local moments, while a
relatively small fraction of the doped carriers has
three-dimensionally extended wave-functions in the crystalline
parts of the sample.~\cite{epsteinprl} Since, in the doped
polyaniline the charge and the spins are associated with the same
carrier,~\cite{mizo} magnetic susceptiblity and ESR measurements
can provide detailed understanding about the microscopic nature of
these carriers. We report here a detailed study of the nature of
charge carriers in a recently reported conducting
PANI,~\cite{sarma} doped with BF$_3$, which has so far been
characterized only by spectroscopic methods. Magnetic
susceptiblity and temperature dependence of ESR linewidth in
conjunction with x-ray diffraction results suggest interesting
distinctive features of this compound, including an order of
magnitude increase in the intrinsic conductivity compared to all
other known conducting PANI systems.

The sample was prepared~\cite{sarma} by doping BF$_3$ into the
emeraldine base, synthesized by chemical oxidation of aniline
using (NH$_4$)$_2$S$_2$O$_8$ at 0-5$^\circ$C.~\cite{smv18} Magnetic susceptiblity
measurement was carried out with powdered sample. Since adsorbed
molecular oxygen induces ESR line broadening, the powder sample was
sealed in a quartz tube under He for those experiments.

In Fig.~\ref{xrd}, we show the x-ray diffraction (XRD) patterns
from undoped PANI and PANI doped with BF$_3$ in comparison to that
from HCl doped PANI, adapted from Ref.~6. The XRD of PANI clearly
suggests that the system is amorphous, in agreement with the previous
result.~\cite{jozefowicz} The appearance of a large number of
sharp diffraction peaks upon doping with HCl, suggests significant
crystallization of PANI on protonation. In contrast, BF$_3$-doped
PANI exhibits an almost identical XRD as the undoped, amorphous
PANI, except for the emergence of a weak intensity peak at
$2\theta~\approx$ 24.8$^\circ$. We find that the position of this
peak matches with the most intense peak in the XRD of
HCl-PANI. More significantly, we find that the intensity of this
peak grows rapidly with progressive time of exposure of the sample
to atmosphere during the recording of XRD, as shown in the inset
to Fig.~\ref{xrd} in the form
of repeated XRD scans of the relevant region. In view of the
extreme sensitivity of BF$_3$ to moisture, forming HF upon
hydrolysis, this extraneous peak in XRD of
BF$_3$-doped PANI is easily understood in terms of hydrolyzed
BF$_3$ reacting with PANI and giving rise to small crystalline
domains of protonated PANI. By minimizing the time of exposure during the
measurement, the intensity of this peak could be kept at a low
level, though it could never be eliminated completely. It is important to
note that the conductivity of the sample was found to decrease
with increasing exposure to the atmospheric conditions,
concomitant with the increase in the intensity of the spurious
peak at 24.8$^\circ$. These observations clearly suggest that
in sharp contrast to proton-doped PANI, BF$_3$-doped samples
remain amorphous as the undoped PANI. Considering that the
intrinsic conductivities of BF$_3$-doped samples are orders of
magnitude higher than that of the protonated samples, as discussed
later, this observation puts the present system in a unique class
of amorphous and yet highly conducting polymers.

We show the susceptibility as a function of temperature in Fig.~2.
The experimentally obtained susceptibility value has
been normalized to a two ring repeat unit and finally corrected
for the diamagnetic contributions of the atomic cores
($\chi_{core}$) using Pascal constants.~\cite{pascal} The
calculated $\chi_{core}$ value for BF$_3$ is 193.6$\times10^{-6}$
emu~mol$^{-1}$. The susceptibility of doped polyaniline has been
described~\cite{epstein87} by $\chi~=~\chi_P~+~C/T$, where, $\chi_P$
is the temperature independent Pauli paramagnetic
contribution from the delocalized charge carriers and $C$ is the
Curie constant arising from the trapped or localized spins in the
polymer. $\chi_P$ can in turn be expressed in terms of the average density of
states at the Fermi level, $N(E_F)$, by the relation $\chi_P~=~\mu_B^2N(E_F)$,
where $\mu_B$ is the Bohr magneton. Thus, $\chi_P$ can provide a good
estimate of the intrinsic metallic conductivity of the sample, which the
bulk resistivity measurement often fails to project, since it depends strongly
on the intergrain connectivity in the sample. For instance, the
$N(E_F)$ value for CSA-PANI~\cite{csachi}, calculated from the $\chi_P$ value is
0.7~states~eV$^{-1}$~(2~rings)$^{-1}$, while that for the HCl-doped
sample is 1.6~states~eV$^{-1}$~(2~rings)$^{-1}$.~\cite{PRBv49} This is
consistent with the fact that the room temperature conductivity of CSA doped
PANI in the pressed pellet form is 0.047~S~cm$^{-1}$,~\cite{csamacro}
which is lower than that of powdered HCl-PANI, though PANI doped
with CSA has the highest bulk conductivity among the family of doped
polyanilines, in thin film form.~\cite{csaprb}

Fig.~\ref{susc}a shows the susceptibility as a function of
temperature ($\chi$~\textit{vs.}~$T$ plot), indicating a nearly
temperature independent Pauli-like susceptiblity in the high
temperature region. The rapid increase of susceptibility with
decreasing $T$ follows a typical Curie behavior, as demonstrated
by the $\chi$~\textit{vs.}~$1/T$ plot in the same figure. This
plot shows the expected linear behavior in the low temperature
regime. The values of $C$, $\chi_P$ and $N(E_F)$ estimated from
the best fit of the susceptibility data are
$3.7\times10^{-3}$~emu~K~mol$^{-1}$,
$380\times10^{-6}$~emu~mol$^{-1}$ and
11.8~states~eV$^{-1}$~(2~rings)$^{-1}$, respectively. Thus, the
calculated density of states at the Fermi level is considerably
higher than that for any other conducting polyaniline system
known, typical values reported~\cite{epsteinprl} so far for
various protonated samples being in the range of
1.6~states~eV$^{-1}$~(2~rings)$^{-1}$. Such a higher concentration
of delocalized charge carriers in BF$_3$ doped PANI suggests a
very high intrinsic conductivity in this system. Also, a high
$\chi_P$ value in predominantly amorphous BF$_3$-doped PANI is in
sharp contrast with the previous results on HCl doped
PANI,~\cite{jozefowicz} which show that the delocalized carriers
and therefore, $\chi_P$ are mainly associated with the crystalline
domains in doped PANI. It is possible that the absence of any
counterions in BF$_3$-PANI, as opposed to the protonated PANI,
may be responsible for these qualitative changes.

Though we have analyzed the magnetic data in terms of $\chi~=~\chi_P~+~C/T$,
close inspection of the temperature dependence suggests interesting
deviations from this behavior at~$\sim$~200~K. For systems obeying the
above equation, $\chi$$T$ is linear in temperature. However, $\chi$$T$ for
the BF$_3$ doped PANI shown in Fig.~\ref{susc}b exhibits an increase in
the slope above 210 K, indicating a slight increase in
$\chi_p$ and a simultaneous decrease in $C$. While this increase in
Pauli susceptibility component is not evident in the $\chi$ \textit{vs.}
$T$ plot shown in Fig.~\ref{susc}a, $ln~\chi$ \textit{vs.} $ln~T$ plot in
Fig.~\ref{susc}b accentuates a small but gradual increase in the free carrier
susceptibility above about 200~K. This enhancement of $\chi_P$ at the cost
of $C$ with increasing $T$ suggests that these changes are possibly
a consequence of the thermal activation of the localized spins into
free carriers, indicating that the trapping potential is in the scale of
the thermal energy in this system.

We find that the ESR signals from BF$_3$-doped PANI have homogenously
broadened Lorentzian
shape, which suggests a three dimensional delocalization of the charge
carriers~\cite{eprlor} as well as a negligible contribution from the
hyperfine and dipolar interactions.~\cite{PRBv45} The ESR linewidth
of a sample is determined by various relaxation
processes that influence the lifetime of the spin carriers.
We present the temperature dependence of the peak-to-peak
linewidth of the ESR signal ($\Delta$$H_{pp}$) for BF$_3$-PANI
in 4-300~K range in Fig.~\ref{epr}. Although the
lineshape remained essentially Lorentzian in the entire temperature
range, the linewidth shows different variations in three different
temperature regimes, marked in the figure as A, B and C.
In systems where the scattering between the localized
and the delocalized spins determines the relaxation rates, the relaxation
time and therefore the linewidth are given by the Korringa relation,~\cite{korringa}
$\Delta$$H_{pp}$~$\propto$~$N^2(E_F)T$. Between 50 K and 230 K (region B),
we indeed find the linewidth to increase linearly with T. Since the slope is
directly proportional to $N^2(E_F)$, we compare the slope in our sample
with that of a previously published result on HCl doped PANI films,~\cite{PRBv45}
for which the $N(E_F)$ is known to be 1.6~states~eV$^{-1}$~(2-rings)$^{-1}$. From the
ratio of the slopes, the calculated $N(E_F)$ for BF$_3$-PANI is
15.2~states~eV$^{-1}$~(2-rings)$^{-1}$; this value matches well with that
obtained from the susceptibility results, once again establishing an
order of magnitude improvement in the intrinsic conductivity.
At further lower temperatures (region A), the linewidth is essentially
temperature independent. This is possibly due to a partial localization of
the mobile spins, thereby increasing the concentration of fixed spins and
compensating the thermally induced changes in the ESR linewidth.
More pronounced effect has been observed~\cite{PRBv45,prb98} in the case of
powdered HCl doped PANI, where the ESR linewidth actually begins to increase
with decreasing temperature below 190 K. In more
ordered samples like stretched films of HCl-doped PANI, the ESR linewidth
at lower temperatures becomes temperature independent,~\cite{prb98}
as observed here. It is therefore quite intriguing to note that despite
being amorphous, the effect of localization is so weak in the case of
BF$_3$-PANI. It would appear that the
energy difference between the chemical potential and the mobility edge
in BF$_3$-doped PANI is very small, though the system is extremely
disordered. At higher temperatures (region C), we observe a sharp
decrease in the ESR linewidth. Since the localized spins are primarily
responsible for the relaxation process in ESR, this observation suggests
that the number of localized spins decreases at these higher temperatures.
This is consistent with the conclusions based on the magnetic susceptibility
data (Fig. 2b) that the delocalized spins increasingly become mobile
above $\sim$ 210~K, leading to an increase of the $\chi_P$ component at the
expense of the $C$ value.

In summary, we report an interesting class of conducting
polyaniline doped with electron deficient boron trifluoride. The
intrinsic conductivity is at least an order of magnitude higher
compared to all other polyaniline samples. This is remarkable in
view of the fact that the polymer remains amorphous even on
doping, representing a highly conducting amorphous state, at least
above~$\sim$~50~K. This is in sharp contrast to proton doped PANI
where the conductivity is associated with the crystalline domains.
The energy scales associated with the trap potential for
localizing spins was found to be small, in order of the thermal
energy scale. These distinctive behaviors suggest that the present
sample belongs to an interesting class of amorphous and yet highly
conducting polymers.

We are grateful to the Council for Scientific and Industrial
Research (CSIR) for providing the financial support.

\newpage
\begin {figure}
\caption{\label{xrd} The x-ray diffraction patterns of undoped,
BF$_3$-doped and 50$\%$ HCl doped PANI (reproduced from Ref.~6). Inset
shows the progressive gain in the intensity of 24.8$^\circ$ peak
upon exposure to atmosphere.}
\end{figure}

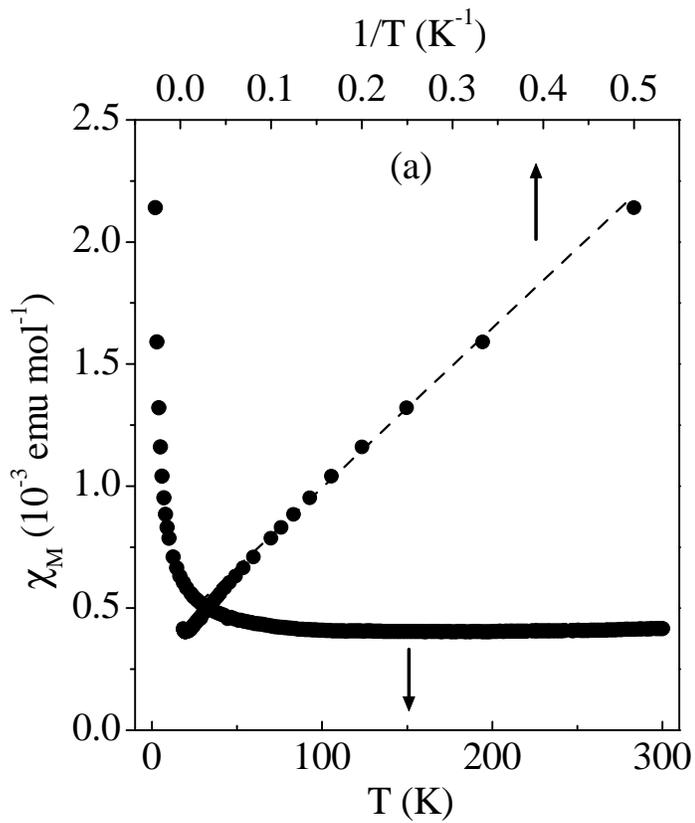
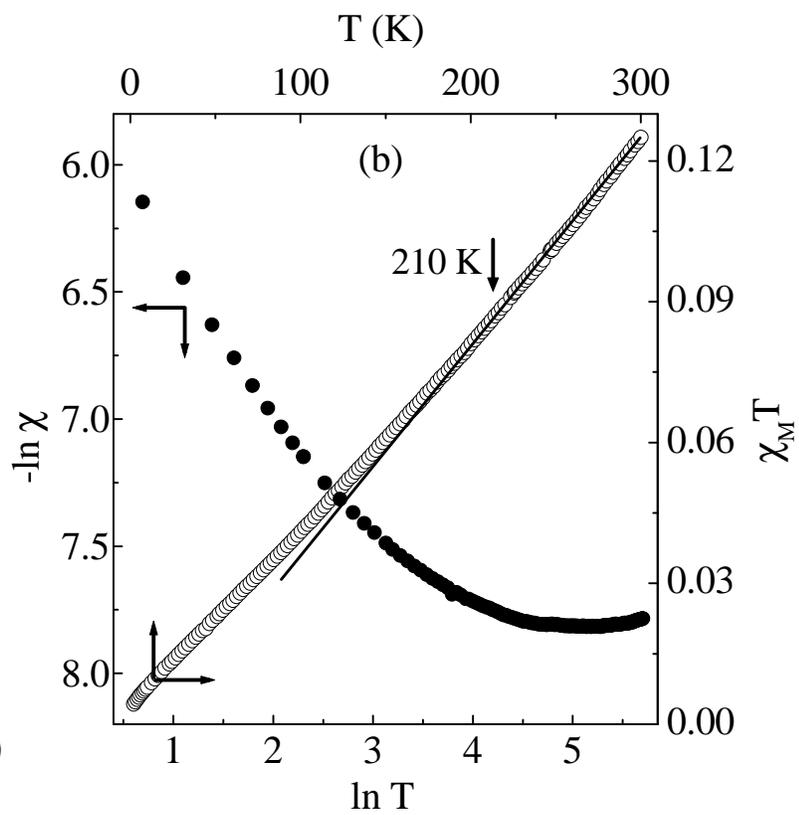
\begin{figure}
\caption{\label{susc} (a): Variation of d.c. magnetic susceptibility
with temperature and inverse temperature. The dashed line is the fit
obtained using the equation, $\chi = \chi_P + C/T$. (b): Plots of $\chi$$T$
vs $T$. The solid line is a guide to the eye. $ln~\chi$ \textit{vs} $ln~T$ suggesting
spin delocalization near 210 K.}
\end{figure}

\begin{figure}
\caption{\label{epr} Plot of peak to peak linewidth of
the ESR signal \textit{vs} $T$ for BF$_3$ doped PANI. The solid
line indicates the linear $T$ dependence in the intermediate regime.}
\end{figure}


\newpage

\begin{thebibliography}{99}

\bibitem[a)]{jnc} Also at Jawaharlal Nehru Centre for
Advanced Scientific Research, Bangalore, India;
electronic mail: sarma@sscu.iisc.ernet.in

\bibitem{hclSM87} A. J. Epstein, J. M. Ginder, F. Zuo, R. W. Bigelow, H. S. Woo,
D. B. Tanner, A. F. Richter, W. S. Huang and A. G. MacDiarmid, Synth. Met. {\bf18},
303, (1987).

\bibitem{epsteinprl} Z. H. Wang, C. Li, E. M. Scherr, A. G. MacDiarmid, and A. J.
Epstein, Phys. Rev. Lett. {\bf66}, 1745 (1991).

\bibitem{mizo} K. Mizoguchi, M. Nechtschein, J. P. Travers, and C. Menardo, Phys. Rev.
Lett. {\bf63}, 66 (1989).

\bibitem{sarma} D. Chaudhuri, A. Kumar, I. Rudra, and D. D. Sarma,
Adv. Mater. {\bf13}, 1548 (2001).

\bibitem{smv18} A. G. MacDiarmid, J. C. Chiang, A. F. Richter, and A. J. Epstein,
Synth. Met. {\bf18}, 285 (1987).

\bibitem{jozefowicz} M. E. Jozefowicz, R. Laversanne, H. H. S. Javadi,
A. J. Epstein, J. P. Pouget, X. Tang, and A. G. MacDiarmid, Phys.
Rev. B {\bf39}, 12958 (1989).

\bibitem{pascal} J. H. Van Vleck, \textit{The Thoery of Electric and Magnetic
Susceptibilities} (Oxford University Press, 1966)

\bibitem{epstein87} J. M. Ginder, A. F. Richter, A. G. MacDiarmid,
and A. J. Epstein, Solid State Commun. {\bf63}, 97 (1987).

\bibitem{csachi} Y. Cao, and A. J. Heeger, Synth. Met. {\bf52}, 193 (1992).

\bibitem{PRBv49} N. S. Sariciftci, A. J. Heeger, and Y. Cao, Phys. Rev. B {\bf49}, 5988
(1994).

\bibitem{csamacro} J. Stejskal, I. Sapurina, M. Trchova, J. Prokes, I. Krivka, and
E. Tobolkova, Macromolecules {\bf31}, 2218 (1998).

\bibitem{csaprb} M. Reghu, Y. Cao, D. Moses, and A. J. Heeger, Phys. Rev. B
{\bf47}, 1758 (1993).

\bibitem{eprlor} J. Joo, V. N. Prigodin, Y. G. Min, A. G. MacDiarmid, and
A. J. Epstein, Phys. Rev. B. {\bf50}, 12226 (1994).

\bibitem{PRBv45} Z. H. Wang, E. M. Scherr, A. G. MacDiarmid, and A. J. Epstein,
Phys. Rev. B {\bf45}, 4190 (1992).

\bibitem{korringa} S. E. Barnes, Adv. Phys. {\bf30}, 801 (1981); J. Korringa, Physica {\bf},
601 (1950).

\bibitem{prb98} J. Joo, S. M. Long, J. P. Pouget, E. J. Oh, A. G. MacDiarmid, and
A. J. Epstein, Phys. Rev. B {\bf57}, 9567 (1998).

\end{thebibliography}
\end{document}